\documentclass[a4paper,11pt]{article}
\usepackage{jinstpub}
\usepackage{lineno}
\usepackage[table]{xcolor}

\title{\boldmath FGPA-based advanced muon trigger algorithms for long-lived particle detection at the High-Luminosity LHC}

\author[1]{D. Estrada,\note{Corresponding author.}}
\author{J. Prado,}
\author{P. Leguina,}
\author{A. Cardini,}
\author{E. Aller,}
\author{and S. Folgueras}
\author{on behalf of the CMS Collaboration\footnote{Copyright 2026 CERN for the benefit of the CMS Collaboration. Reproduction of this article or parts of it is allowed as specified in the CC-BY-4.0 license.}.}
\affiliation{University of Oviedo,\\C. San Francisco 3, Oviedo, Spain}
\affiliation{Instituto de Ciencias y Tecnologías Espaciales de Asturias (ICTEA),\\ C. de la Independencia 13, Oviedo, Spain}

\emailAdd{destrada@cern.ch}

\abstract{The CMS Level-1 trigger must reconstruct muons in real time under stringent latency and hardware resource constraints. Conventional algorithms lose efficiency for high-energy muons undergoing radiative energy losses and can be suboptimal for displaced trajectories, limiting sensitivity to exotic signatures such as long-lived particles (LLPs). The Phase-2 upgrade introduces advanced FPGAs that enable more sophisticated reconstruction algorithms. This work presents two complementary developments addressing these challenges. In the barrel region, a dedicated algorithm detects showers from Drift Tube hit patterns at the first processing stage. At the subsequent filtering stage, shower primitives are correlated with conventional trigger primitives, currently targeting radiatively showering high-momentum muons, with LLP signatures as a broader objective. Simulation studies yield approximately $80\%$ shower-classification accuracy and an estimated matching efficiency above $80\%$ for radiating muons. Firmware prototypes demonstrate a shower-detection latency of approximately $20\,\mathrm{ns}$ and sub-bunch-crossing latency for preliminary matching logic. In parallel, a hardware-oriented optimization methodology is investigated for future graph neural network based displaced-muon reconstruction in the barrel--endcap overlap region. Using a GraphSAGE proxy on a benchmark dataset, a workflow combining integer quantization, power-of-two scaling, and data-driven bit-width optimization achieves $(75.0 \pm 1.1)\%$ accuracy and a deterministic latency of $19$ cycles ($52.8\,\mathrm{ns}$ at $360\,\mathrm{MHz}$), with only $20\%$ digital signal processor utilization. These results establish a basis for implementation studies using CMS muon data.}
\keywords{Trigger algorithms, pattern recognition, particle identification methods.}

\begin{document}
\maketitle
\flushbottom

\section{Introduction}
\label{sec:intro}
Identifying long-lived particles (LLPs) at the Large Hadron Collider (LHC) extends the sensitivity of the Compact Muon Solenoid (CMS) experiment~\cite{CMS:2008xjf,CMS:2023gfb} to physics beyond the Standard Model (BSM). Depending on their decay positions and final states, LLPs can produce displaced muons, displaced hadronic jets, or shower-like activity in the muon system. These signatures differ from those of prompt particles and may evade conventional trigger algorithms optimized for standard trajectories~\cite{Folgueras:2025,Syrunyan:2021,Cottin:2023}.

At the High-Luminosity LHC (HL-LHC), the CMS Level-1 trigger (L1T) must process detector data at a bunch-crossing (BX) frequency of $40\,\mathrm{MHz}$ and select events at a maximum rate of $750\,\mathrm{kHz}$ within a latency budget of $12.5\,\mu\mathrm{s}$. The Phase-2 upgrade introduces advanced field-programmable gate arrays (FPGAs) with greater processing capability, enabling more sophisticated reconstruction algorithms within these constraints. For the algorithms considered here, the processing window upstream of the Global Muon Trigger (GMT) is approximately $4\,\mu\mathrm{s}$. Low latency and predictable resource usage are therefore essential design requirements~\cite{CMS:2020,Zabi:2022}.

This work addresses two complementary challenges in the barrel muon system and the barrel--endcap overlap region. In the barrel, showers can arise from unconventional signatures such as LLP decays or from radiative energy losses of high-momentum muons. Both can lead to a more complex muon reconstruction, affecting sensitivity to BSM physics and the efficiency for $\mathrm{TeV}$-scale muons~\cite{Sirunyan:2020,Acevedo:2026}. In the overlap region, the Overlap Muon Track Finder (OMTF) is used for muon track finding. It matches detector information to predefined reference patterns;  however, this methodology restricts its efficiency for displaced muons with non-prompt trajectories.~\cite{Bunkowski:2019,Zabolotny:2016,Leguina:2025,Cardini:2026}.

Two developments are presented to address the above: a barrel shower-tagging algorithm that combines hit-multiplicity detection with geometrical matching to conventional muon trigger primitives (TPs), and a hardware-oriented optimization methodology for Graph Neural Network (GNN) inference for displaced-muon reconstruction in the OMTF. The shower-tagging study includes simulation results and FPGA implementation, while the GNN study evaluates the optimization workflow using a benchmark graph dataset and a GraphSAGE proxy model.

\section{Muon shower tagging in the CMS barrel system}
\label{sec:shower}
Shower-based triggering is already used to select unconventional signatures in the muon system. During Run 3, the CMS High Multiplicity Trigger (HMT) uses hit multiplicities in endcap Cathode Strip Chambers (CSCs) to select events compatible with LLP decays~\cite{Abbott_2026}.

\begin{figure}[ht]
    \centering
    \includegraphics[width=\linewidth]{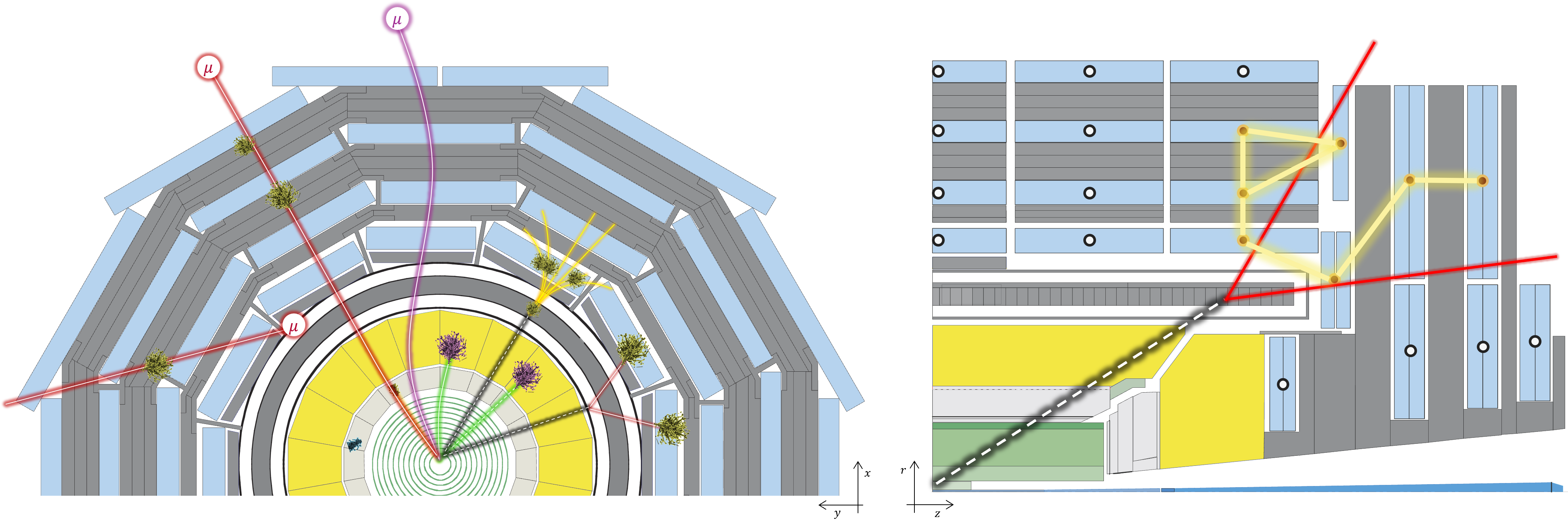}
    \caption{Left: Illustrative examples of shower-like signatures in the CMS muon system, including radiative energy losses of high-energy muons (red) and showers associated with possible LLP decays (yellow). Right: Conceptual graph representation of detector measurements for GNN-based muon reconstruction in the overlap region, nodes represent detector stubs, while edges encode geometrical relationships between compatible stubs.}
    \label{fig:cms-showers-cases+gnn}
\end{figure}

Similarly, in the barrel, showers can originate from LLP decays beyond the hadron calorimeter, including displaced hadronic jets from neutral long-lived scalars, gluinos, or squarks, as well as heavy neutral lepton (HNL) decays~\cite{Syrunyan:2021,Cottin:2023}. Showers can also be produced by high-energy muons, including cosmic muons and muons from heavy BSM vector boson decays such as $Z' \rightarrow \mu\mu$ and $W' \rightarrow \mu\nu$, when they undergo radiative interactions such as bremsstrahlung in the steel flux-return yoke or other detector material (see Fig.~\ref{fig:cms-showers-cases+gnn} left). These showers increase hit multiplicities and can introduce ambiguities in TP generation, degrading conventional muon reconstruction, which has been shown to reduce L1T efficiency. This effect increases with muon momentum, reaching approximately \(10\%\) at \(1\,\mathrm{TeV}\) in the central barrel---CMS region with pseudorapidity ($\eta$) in $[-9, 9]$~\cite{Sirunyan:2020}. Dedicated shower tagging can therefore help identify showering muons and improve sensitivity both for  unconventional signatures and highly energetic muons.

The Phase-2 barrel muon trigger separates local hit reconstruction from multi-station correlation. Its architecture comprises $30$ custom ATCA BMTL1 boards that process Drift Tube (DT) and Resistive Plate Chamber (RPC) data, generating local TPs via the Analytical Method (AM)~\cite{Abbiendi:2023} on AMD/Xilinx Virtex UltraScale+ VU13P FPGAs. Outputs are sent to $12$ Barrel Filter (BF) boards, each receiving data from $10$ BMTL1 boards to combine information across sectors and wheels~\cite{Zabi:2022,CMS:2020}. The BF stage can therefore correlate conventional TPs with shower activity before forwarding candidates to the GMT.

The proposed shower-tagging approach comprises two components. First, shower activity is detected directly from raw DT hits at the BMTL1 stage in parallel with the AM. Second, at the BF stage, the resulting Shower Primitives (SPs) are correlated with conventional TPs to identify signatures of interest. The current implementation focuses on radiative high-momentum muons, while the broader goal includes LLP decays.

The identification algorithm monitors raw DT-hit multiplicities within a $400\,\mathrm{ns}$ window ($16$ BXs, corresponding to the maximum DT electron drift time). When the accumulated hit count exceeds a configurable threshold, the algorithm produces an SP containing the first-hit BX, the minimum and maximum wire identifiers, and a wire hit-count profile vector. This compact representation preserves key timing and spatial information about the shower without transferring the full hit list downstream~\cite{Prado:2026,Acevedo:2026}.

At the BF, for radiative-muon tagging, the algorithm evaluates TPs from neighbouring sectors for each SP and applies timing cuts to reject incompatible candidates. Geometrically, the shower is modeled as a segment between its extreme wire positions, and each TP as a straight-line trajectory. If a TP trajectory intersects the shower segment within the timing window, it is flagged as shower-associated for downstream reconstruction~\cite{Acevedo:2026}. This approach reduces the geometrical test to operations suitable for FPGA implementation.

Simulation studies show an overall shower-identification accuracy of approximately $80\%$, prioritizing minimal missed shower events, with an estimated local TP rate about ten times lower than that of the AM algorithm. The geometrical matching algorithm achieves an estimated efficiency above $80\%$ for tagging showering high-momentum muons.

The initial FPGA implementation of shower identification uses approximately $2.5\,\mathrm{k}$ lookup tables (LUTs) and $2.5\,\mathrm{k}$ flip-flops (FFs), with a latency of $20\,\mathrm{ns}$. For the BF matching logic, a firmware prototype processes up to $48$ TP--SP pairs per clock cycle at $480\,\mathrm{MHz}$, achieving sub-BX latency with approximately $4.3\,\mathrm{k}$ LUTs and $5.6\,\mathrm{k}$ FFs. The current version of the last covers $3$ sectors in a single wheel; scaling it to the full BF coverage remains necessary.

A preliminary parallel development uses the SP density profile as input to a deep neural network (DNN) to distinguish hadron-like showers from background activity. This approach could support the selection of displaced hadronic LLP decays in the barrel. The baseline shower-generation design was integrated into CMS during a pre-shutdown slice test, and its performance with cosmic-ray data is being evaluated.

\section{Hardware-oriented optimization of GNNs for displaced-muon reconstruction}
\label{sec:gnn}
In the barrel--endcap transition region ($0.83 < |\eta| < 1.24$), the OMTF combines DT, CSC, and RPC stubs and matches them to predefined reference patterns using a naive Bayes classifier~\cite{Bunkowski:2019,Zabolotny:2016}. Although effective for prompt muons, this approach is less suitable for displaced trajectories from LLP decays. GNNs offer an alternative by exploiting spatial relationships between detector measurements~\cite{Leguina:2025,Cardini:2026}.

In a graph representation (see Fig.~\ref{fig:cms-showers-cases+gnn} right), nodes encode detector stubs, including layer information and their $\eta$ and $\phi$ coordinates. Edges connect geometrically compatible stubs encoding relative angular differences ($\Delta\eta$ and $\Delta\phi$)~\cite{Leguina:2025}. The reconstruction task would be to identify compatible stubs, reject combinations arising from pile-up, and infer muon trajectory parameters.

GraphSAGE is an inductive architecture that updates node representations by aggregating information from neighbouring nodes~\cite{Hamilton:2017}. This message-passing structure provides a basis for combining local measurements across stations. However, its irregular data dependencies and arithmetic intensity challenge FPGA implementation under the strict latency and resource limits of the L1T~\cite{Iiyama:2020,Cardini:2026}.

This study separates the assessment of hardware feasibility from the development of a physics-specific model. A two-layer GraphSAGE firmware proxy was evaluated on the Cora citation dataset~\cite{Sen:2008}, in which nodes represent publications and edges represent citations. Fixed-size subgraphs containing eight nodes and extending to two hops are extracted to match the target application size, and original $1433$-dimensional node features are projected off-chip to $16$ features (this is excluded from inference-latency measurement). The proxy provides a controlled environment for assessing message-passing timing and resource usage, but does not evaluate CMS physics performance~\cite{Cardini:2026}.

While conventional GraphSAGE combines aggregated neighbour information with a self-term from the root node~\cite{Hamilton:2017}, the firmware proxy omits this contribution. This simplification removes one weight matrix, reduces the number of multiply-accumulate (MAC) operations, eliminates intermediate concatenation buffers, and simplifies fixed-size loops for full array unrolling in high-level synthesis (HLS)~\cite{Cardini:2026}.

Starting from a PyTorch floating-point model, the optimization workflow followed comprises fixed-size graph extraction, integer-only post-training quantization (PTQ), power-of-two (PO2) scale approximation, data-driven bit-width optimization, and bit-exact validation against a Python integer reference~\cite{Cardini:2026}. The discussion below focuses on PO2 scaling and bit-width optimization concept before presenting the results.

In standard INT8 quantization, a neuron's accumulated result ($acc$) is rescaled before the next layer to prevent data overflow~\cite{Jacob:2018}. Mathematically, this standard rescaling is expressed as:
\begin{equation*}
    q_{\text{out}} =
    \operatorname{clamp}\left((acc \cdot \alpha) \gg M\right),
\end{equation*}
where $\alpha$ is a fixed-point scale multiplier and $M$ is a binary-point bit shift. On FPGA, general fixed-point multiplication requires dedicated digital signal processor (DSP) blocks and increases resource usage. The INT8-PO2 optimization approximates each scale factor by the nearest power of two ($\alpha \approx 2^p$)~\cite{Cardini:2026}, reducing rescaling to a simple bit shift:
\begin{equation*}
    q_{\text{out}} =
    \operatorname{clamp}(acc \gg k),
    \qquad k = M-p.
\end{equation*}
This removes general rescaling multiplications, leaving matrix multiplication as the main consumer of DSP resources.

Data-driven bit-width optimization further reduces resource usage. Rather than using uniform precision or worst-case estimates, the method analyses intermediate signal distributions and assigns each signal the minimum bit width needed to prevent overflow, with an additional safety margin. The resulting widths are used to generate layer-specific HLS parameters~\cite{Cardini:2026}.

\begin{table}[htb]
    \centering
    \resizebox{\linewidth}{!}{%
    \begin{tabular}{lcccccc}
        \hline
        Configuration & Accuracy (\%) & Latency (cycles) & Latency (ns) & DSP (\%) & FF (\%) & LUT (\%) \\
        \hline
        Float32          & $78.0 \pm 0.8$ & 603 & 1675.0 & 283 & 128 & 123 \\
        Fixed-PTQ        & $75.7 \pm 1.0$ & 77  & 213.9  & 56  & 7   & 6   \\
        INT8-PO2         & $75.0 \pm 1.2$ & 28  & 77.8   & 43  & 7   & 10  \\
        QAT              & $76.8 \pm 0.9$ & 55  & 152.8  & 55  & 16  & 16  \\
        \rowcolor{gray!25} INT8-PO2 Opt. & $75.0 \pm 1.1$ & 19 & 52.8 & 20 & 6 & 27 \\
        \hline
    \end{tabular}%
    }
    \caption{Accuracy (mean $\pm$ standard deviation over five random seeds), latency as number of cycles or in ns assuming a clock frequency of 360 MHz, and XCVU13P FPGA resource utilization ($>100\%$ exceeds the capacity of a single device) at $360\,\mathrm{MHz}$. Adapted from~\cite{Cardini:2026}.}
    \label{tab:gnn-results}
\end{table}

Table~\ref{tab:gnn-results} summarizes the FPGA synthesis results. Combining INT8-PO2 quantization with data-driven bit-width optimization reduces inference latency to $19$ clock cycles ($52.8\,\mathrm{ns}$ at $360\,\mathrm{MHz}$), using $20\%$ of DSPs, $6\%$ of FFs, and $27\%$ of LUTs. The optimized configuration achieves $(75.0 \pm 1.1)\%$ accuracy, compared with $(78.0 \pm 0.8)\%$ for the floating-point reference. The Python integer reference and HLS C-simulation also show bit-exact agreement~\cite{Cardini:2026}.

\section{Conclusions and perspectives}
\label{sec:conclusions}
Two complementary approaches for identifying unconventional muon signatures in the CMS Phase-2 L1T have been investigated within the latency and FPGA resource limits upstream of the GMT.

In the barrel, a firmware shower-identification algorithm generates SPs with low FPGA resource usage and sub-BX latency. Simulation studies indicate promising performance for shower detection and the tagging of radiatively showering high-momentum muons. The baseline design was already integrated into CMS during a pre-shutdown slice test, and evaluation with cosmic-ray data is ongoing.

For the OMTF, an FPGA-oriented optimization workflow has been developed to guide GNN implementation in hardware. It combines integer quantization, PO2 scaling, and data-driven bit-width fine-tuning, achieving a deterministic latency of $19$ cycles and low DSP usage while maintaining accuracy close to the floating-point reference on a GraphSAGE proxy model. Validation with physics-driven displaced-muon datasets and implementation in the target FPGA remain necessary next steps.

Future barrel developments include extending the matching logic to full detector coverage, incorporating also longitudinal $\theta$ primitives, and using SPs to identify signatures of LLP displaced vertices. Future GNN studies will apply the workflow to realistic CMS muon data and explore architectures for track association, charge identification, and transverse-momentum reconstruction.

\section*{Acknowledgements}

This work is supported by ERC grant INTREPID, 10.3030/101115353. Funded by the European Union. Views and opinions expressed are, however, those of the author(s) only and do not necessarily reflect those of the European Union or the European Research Council Executive Agency. Neither the European Union nor the granting authority can be held responsible for them.

\bibliographystyle{JHEP}
\bibliography{biblio.bib}

@techreport{CMS:2020,
    author      = "{CMS Collaboration}",
    title       = "{The Phase-2 Upgrade of the CMS L1 Trigger Technical Design Report}",
    institution = "CERN",
    number      = "CERN-LHCC-2020-004, CMS-TDR-021",
    year        = "2020",
    address     = "Geneva, Switzerland",
    url = {https://cds.cern.ch/record/2714892?ln=es},
}

@inproceedings{Zabi:2022,
  author    = "Alexandre Zabi and others",
  title     = "{System Design and Prototyping for the CMS Level-1 Trigger at the High-Luminosity LHC}",
  doi       = "10.1088/1742-6596/2374/1/012090",
  booktitle = "Proceedings of the International Conference on Technology and Instrumentation in Particle Physics {\textemdash} J. Phys.: Conf. Ser. (TIPP 2021)",
  year      = 2022,
  volume    = "2374",
  pages     = "012090"
}

@article{Folgueras:2025,
   author = {Santiago Folgueras},
   title = {{INTREPID}},
   journal={108-111, Available at https://www.europeandissemination.eu/article/intrepid/22858},
   url = {https://www.europeandissemination.eu/article/intrepid/22858},
   year = {2025},
}

@inproceedings{Acevedo:2026,
  author = "Daniel Estrada  and others",
  title = "{Triggering on muon showers in the Barrel Muon Trigger of the CMS experiment for the HL-LHC upgrades}",
  doi = "10.22323/1.485.0577",
  booktitle = "Proceedings of The European Physical Society Conference on High Energy Physics {\textemdash} PoS(EPS-HEP2025)",
  year = 2026,
  volume = "485",
  pages = "577"
}

@inproceedings{Leguina:2025,
    author = "Leguina, Pelayo and others",
    title = "{Using AI on FPGAs for the CMS Overlap Muon Track Finder for the HL-LHC}",
    booktitle = "{2nd European AI for Fundamental Physics Conference}",
    eprint = "arXiv:2509.23347",
    archivePrefix = "arXiv",
    primaryClass = "hep-ex",
    month = "9",
    year = "2025"
}

@article{Sirunyan:2020,
doi = {10.1088/1748-0221/15/02/P02027},
url = {https://doi.org/10.1088/1748-0221/15/02/P02027},
year = {2020},
month = {feb},
publisher = {},
volume = {15},
number = {02},
pages = {P02027},
author = {{CMS Collaboration}},
title = {Performance of the reconstruction and identification of high-momentum muons in proton-proton collisions at $\sqrt{s}=13\text{ }\text{ }\mathrm{TeV}$},
journal = {Journal of Instrumentation}
}

@article{Abbiendi:2023,
    author  = "Abbiendi, G. and others",
    title   = "{The Analytical Method algorithm for trigger primitives generation at the LHC Drift Tubes detector}",
    journal = "Nuclear Instruments and Methods in Physics Research Section A (NIMA)",
    volume  = "1049",
    pages   = "168103",
    year    = "2023",
    doi={https://doi.org/10.1016/j.nima.2023.168103}
}

@inproceedings{Prado:2026,
  author    = "J. Prado and others",
  title     = "{Muon shower tagging in the barrel muon system of the CMS experiment}",
  doi       = "10.1088/1748-0221/21/01/C01050",
  booktitle = "Proceedings of the 26th International Workshop on Radiation Imaging Detectors {\textemdash} JINST (IWORID2025)",
  year      = 2026,
  volume    = "21",
  pages     = "C01050"
}

@inproceedings{Bunkowski:2019,
  author    = "K. Bunkowski and others",
  title     = "{The algorithm of the CMS Level-1 Overlap Muon Track Finder trigger}",
  doi       = "10.1016/j.nima.2018.10.173",
  booktitle = "Proceedings of the 14th Pisa Meeting on Advanced Detectors (Frontier Detectors for Frontier Physics) {\textemdash} Nucl. Instrum. Meth. A",
  year      = 2019,
  volume    = "936",
  pages     = "368--369"
}

@inproceedings{Zabolotny:2016,
  author    = "W. M. Zabolotny and others",
  title     = "{Algorithm and implementation of muon trigger and data transmission system for barrel-endcap overlap region of the CMS detector}",
  doi       = "10.1088/1748-0221/11/03/C03004",
  booktitle = "Proceedings of the Topical Workshop on Electronics for Particle Physics {\textemdash} JINST (TWEPP 2015)",
  year      = 2016,
  volume    = "11",
  pages     = "C03004"
}

@inproceedings{Hamilton:2017,
  author    = "Will Hamilton and others",
  title     = "{Inductive Representation Learning on Large Graphs}",
  booktitle = "Advances in Neural Information Processing Systems {\textemdash} NeurIPS",
  year      = 2017,
  volume    = "30",
  pages     = "1024--1034",
  eprint        = "arXiv:1706.02216",
  archivePrefix = "arXiv",
  primaryClass  = "cs.SI",
  url       = "https://proceedings.neurips.cc/paper\_files/paper/2017/file/5dd9db5e033da9c6fb5ba83c7a7ebea9-Paper.pdf"
}

@article{Iiyama:2020,
    AUTHOR={Iiyama, Yutaro  and Cerminara, Gianluca  and Gupta, Abhijay  and Kieseler, Jan  and Loncar, Vladimir  and Pierini, Maurizio  and Qasim, Shah Rukh  and Rieger, Marcel  and Summers, Sioni  and Van Onsem, Gerrit  and Wozniak, Kinga Anna  and Ngadiuba, Jennifer  and Di Guglielmo, Giuseppe  and Duarte, Javier  and Harris, Philip  and Rankin, Dylan  and Jindariani, Sergo  and Liu, Mia  and Pedro, Kevin  and Tran, Nhan  and Kreinar, Edward  and Wu, Zhenbin },
           
    TITLE={Distance-Weighted Graph Neural Networks on {FPGA}s for Real-Time Particle Reconstruction in High Energy Physics},
    JOURNAL={Frontiers in Big Data},
    VOLUME={Volume 3 - 2020},
    YEAR={2021},
    URL={https://www.frontiersin.org/journals/big-data/articles/10.3389/fdata.2020.598927},
  
DOI={10.3389/fdata.2020.598927},
  
ISSN={2624-909X},
}

@article{Sen:2008,
    author  = "Sen, P. and Namata, G. and Bilgic, M. and Getoor, L. and Gallagher, B. and Eliassi-Rad, T.",
    title   = "{Collective Classification in Network Data}",
    journal = "AI Magazine",
    volume  = "29",
    number  = "3",
    pages   = "93--106",
    year    = "2008",
    doi     = "10.1609/aimag.v29i3.2157"
}

@inproceedings{Jacob:2018,
    author = { Jacob, Benoit and Kligys, Skirmantas and Chen, Bo and Zhu, Menglong and Tang, Matthew and Howard, Andrew and Adam, Hartwig and Kalenichenko, Dmitry },
    booktitle = { 2018 IEEE/CVF Conference on Computer Vision and Pattern Recognition (CVPR) },
    title = {{ Quantization and Training of Neural Networks for Efficient Integer-Arithmetic-Only Inference }},
    year = {2018},
    volume = {},
    ISSN = {},
    pages = {2704-2713},
    doi = {10.1109/CVPR.2018.00286},
    url = {https://doi.ieeecomputersociety.org/10.1109/CVPR.2018.00286},
    publisher = {IEEE Computer Society},
    address = {Los Alamitos, CA, USA},
    month =Jun
}

@article{Syrunyan:2021,
  title         = {Search for long-lived particles using displaced jets in proton-proton collisions at $\sqrt{s}=13\text{ }\text{ }\mathrm{TeV}$},
  author        = {{CMS Collaboration}},
  journal = {Phys. Rev. D},
  volume = {104},
  issue = {1},
  pages = {012015},
  numpages = {38},
  year = {2021},
  month = {Jul},
  publisher = {American Physical Society},
  doi = {10.1103/PhysRevD.104.012015},
  url = {https://link.aps.org/doi/10.1103/PhysRevD.104.012015}
}

@Article{Cottin:2023,
    author={Cottin, Giovanna
    and Helo, Juan Carlos
    and Hirsch, Martin
    and others},
    title={Long-lived heavy neutral leptons with a displaced shower signature at {CMS}},
    journal={Journal of High Energy Physics (JHEP)},
    year={2023},
    volume={02},
    pages={011},
    issn={1029-8479},
    doi={10.1007/JHEP02(2023)011},
    url={https://doi.org/10.1007/JHEP02(2023)011}
}

@inproceedings{Cardini:2026,
  author        = "Andrea Cardini and others",
  title         = "{Integer Quantization of Graph Neural Networks for Real-Time FPGA Track Finding}",
  booktitle     = "Proceedings of the 25th IEEE Real Time Conference (RT2026)",
  year          = 2026,
  eprint        = "arXiv:2609.28144",
  archivePrefix = "arXiv",
  primaryClass  = "hep-ex",
}

@article{Abbott_2026,
doi = {10.1088/1748-0221/21/07/P07040},
url = {https://doi.org/10.1088/1748-0221/21/07/P07040},
year = {2026},
month = {jul},
publisher = {IOP Publishing},
volume = {21},
number = {07},
pages = {P07040},
author = {Abbott, M. and others},
title = {High Multiplicity Trigger for long-lived particles in {CMS} detector},
journal = {Journal of Instrumentation}
}

@article{CMS:2008xjf,
    author = "Chatrchyan, S. and others",
    collaboration = "CMS",
    title = "{The {CMS} Experiment at the CERN LHC}",
    doi = "10.1088/1748-0221/3/08/S08004",
    journal = {Journal of Instrumentation (JINST)},
    volume = "3",
    pages = "S08004",
    year = "2008"
}

@article{CMS:2023gfb,
doi = {10.1088/1748-0221/19/05/P05064},
url = {https://doi.org/10.1088/1748-0221/19/05/P05064},
year = {2024},
month = {may},
publisher = {IOP Publishing},
volume = {19},
number = {05},
pages = {P05064},
author = {{CMS collaboration}},
title = {Development of the {CMS} detector for the {CERN LHC Run3}},
journal = {Journal of Instrumentation (JINST)},
}

\end{document}